\documentclass{PoS}
\newcommand{\pt}{\ensuremath{p_{T}}}

\newcommand{\zg}{\ensuremath{z_g}~}
\newcommand{\rg}{\ensuremath{R_g}~}
\usepackage{url}

\title{Jet sub-structure and parton shower evolution in p+p and Au+Au collisions at STAR}

\ShortTitle{Jet sub-structure at STAR}

\author{\speaker{Raghav Kunnawalkam Elayavalli for the STAR Collaboration}\\
        Wayne State University, Detroit MI 48201\\
        E-mail: \email{raghavke@wayne.edu}}

\abstract{
Recent measurements of jet structure modifications at RHIC and LHC highlight the importance of differential measurements to study the nature of jet quenching. Since these jet structure observables are intimately dependent on parton evolution in both the angular and energy scales, measurements are needed to disentangle these two scales in order to probe the medium at different length scales to study its characteristic properties such as the coherence length. To that effect, the STAR collaboration presents fully unfolded results of jet sub-structure observables designed to extract fundamental quantities related to the parton shower via the SoftDrop shared momentum fraction ($z_{g}$), the groomed jet radius ($R_{g}$) and the jet Mass ($M$) in p+p collisions at $\sqrt{s} = $ 200 GeV as a function of jet transverse momenta. We also showcase the first measurement of iterative softdrop groomed $z_{g}$ and $R_{g}$ for first, second and third splits with an initiator prong transverse momenta ranging from 20-25 GeV. In comparing the un-corrected data to our simulation, we are able to look at snapshots of the jet clustering history leading towards an understand of the time evolution of the parton shower. Having established the p+p baseline, we present the first measurement of the jet's inherent angular structure in Au+Au collisions at $\sqrt{s_{\mathrm{NN}}} = $ 200 GeV via an experimentally robust observable related to the SoftDrop $R_{g}$: the opening angle between the two leading sub-jets ($\theta_{SJ}$). In Au+Au collisions at STAR, we utilize a specific di-jet selection as introduced in our previous momentum imbalance ($A_{J}$) measurement and measure both the $A_{J}$ and the recoil jet spectra differentially as a function of the angular classes based on the $\theta_{SJ}$ observable. With such measurements, we probe the medium response to jets at a particular resolution scale and find no significant differences in quenching for jets of different angular scales as given by $\theta_{SJ}$.}

\FullConference{13th International Workshop on High-pT Physics in the RHIC/LHC era\\
		19 - 22 March 2019\\
		Knoxville, Tennessee, USA}

\begin{document}

\section{Introduction}
Relativistic ion collisions produce copious amounts of jets due to the hard scatterings between quarks and gluons of the colliding nuclei. 
Recent measurements at both RHIC and LHC along with theoretical advancements have shown the importance of studying and measuring the properties of these jets both in p+p and in heavy ion collisions (review of jet studies can be found here~\cite{expreview}). 
There are two natural scales that characterize a jet and its evolution: the momentum and the angular scales. First generation jet measurements have measured jet quenching in an integrative manner, for example via the momentum asymmetry in di-jet events and have further extensively studied the momentum dependence via nuclear modification factors and fragmentation functions. Jet-medium interaction could further be dependent on the resolution scale or the coherence length of the medium which perceives the jet as a singular radiating object or a multi-prong object~\cite{cohlength}. 

\section{Jet sub-structure in p+p Collisions}

In order to differentially study energy loss in the QGP medium, the jet structure in vacuum has to be first understood at both momentum and virtuality scales related to the DGLAP splitting functions~\cite{dglap} that govern parton evolution. SoftDrop~\cite{SoftDrop} is an algorithm with which one can extract such scales experimentally utilizing its procedure of walking backwards in the Cambridge/Aachen clustering tree until two sub-jets satisfy $z = \frac{min(p_{T,1}, p_{T,2})}{p_{T,1} + p_{T,2}} > z_{cut} (\Delta R/R)^{\beta}$ where \zg and \rg are the $z$ and $\Delta R$ upon termination of the algorithm with $z_{cut} = 0.1$ and $\beta = 0$. It was shown that for such choices of $z_{cut}, \beta$ the SoftDrop $z_{g}$ distribution converges to the vacuum DGLAP splitting functions for $z>z_{cut}$ in a ``Sudakov-safe'' manner~\cite{Larkoski:2015lea}. The invariant jet mass is also measured as a function of jet $p_{T}$ since it is dependent on the virtuality scale set by the parton shower. 

\begin{figure}[h] 
   \centering
   \includegraphics[width=0.98\textwidth]{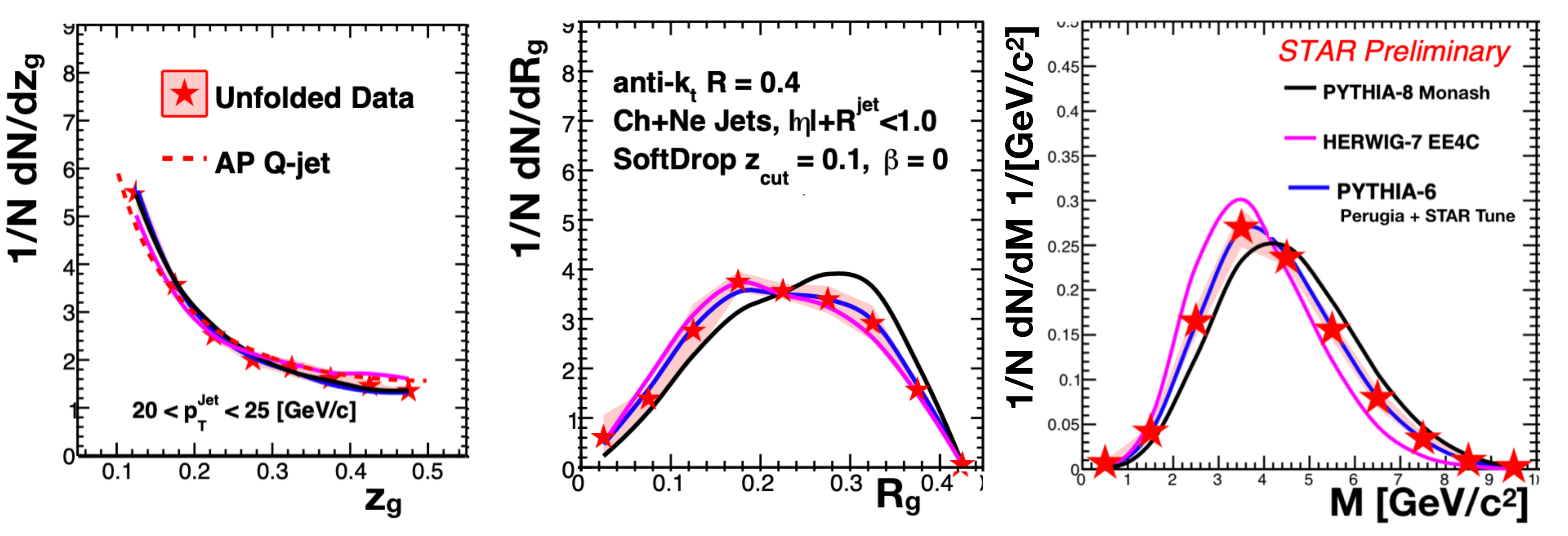} 
   \caption{Fully unfolded measurement of the SoftDrop groomed sub-jet shared momentum fraction ($z_{g}$), the groomed jet radius ($R_{g}$) and the jet mass in p+p collisions at $\sqrt{s} =$ 200 GeV for jets with $20 < p_{T} < 25$ GeV/$c$.}
   \label{fig:ppzgrgm}
\end{figure}

The p+p data for the jet sub-structure measurements were collected with the STAR detector~\cite{star} during the 2012 run at $\sqrt{s_{\mathrm{NN}}} = 200$ GeV.  Jets are reconstructed from charged tracks in the Time Projection Chamber (TPC) and energy depositions in the Barrel ElectroMagnetic Calorimeter (BEMC) using the anti-$k_{t}$ algorithm as implemented in the FastJet package~\cite{FastJet}, hereafter referenced as Ch+Ne jets. 
For additional details regarding event/track/tower quality selections, please refer to~\cite{starprl, nicktalk, raghavHP18}. 

The $z_{g}$, $R_{g}$ and $M$ distributions versus jet $p_{T}$ are corrected using two-dimensional bayesian unfolding as implemented in the RooUnfold package~\cite{roounfold, dgostini} with four iterations. The response matrix is created from a PYTHIA-6 (Perugia Tune, slightly adapted to STAR data)~\cite{pythia6, starpythiatune} prior and a GEANT-3 simulation of the STAR detector. The systematic uncertainties for data are taken as a quadrature sum resulting from the following sources: tracking efficiency ($4\%$), tower gain calibration ($3.8\%$), hadronic correction to the tower energy scale (described in ~\cite{starprl}) and unfolding-related sources including varying the iteration parameter from 2 to 6 and the prior in the response matrix.  

Figure~\ref{fig:ppzgrgm} shows the fully unfolded \zg (left), \rg (middle) and $M$ (right) distributions for jets with $ 20 < p_{T} < 25 $ GeV/$c$, respectively. The STAR data are shown in the red filled star markers with the red shaded region corresponding to the overall systematic uncertainty. Leading order Monte Carlo (MC) generators such as PYTHIA-6 Perugia tune, PYTHIA-8 Monash~\cite{pythia8}, and Herwig-7 EE4C UE tune~\cite{herwig7} in the blue, black and magenta lines are also plotted for comparison to the data. For the \zg observable, we also provide the symmetrized DGLAP splitting functions (noted as AP Q-Jet in the figure) at leading order in the red dashed lines for quark jets (with the splitting being similar for quark- and gluon-initiated jets). All the models studied reproduce the general trends seen in the data, particularly the dependence on the jet momenta leading to a steeper \zg distribution and a narrower $R_{g}$. As the jet mass is related to both the momentum and angular scales, the differences we see in both \zg and $R_{g}$ carry over onto the mass with PYTHIA-8 and HERWIG-7 predicting larger and smaller masses respectively.   

\section{Iterative Jet-Substructure in p+p Collisions}

\begin{figure}[h] 
   \centering
   \includegraphics[width=0.98\textwidth]{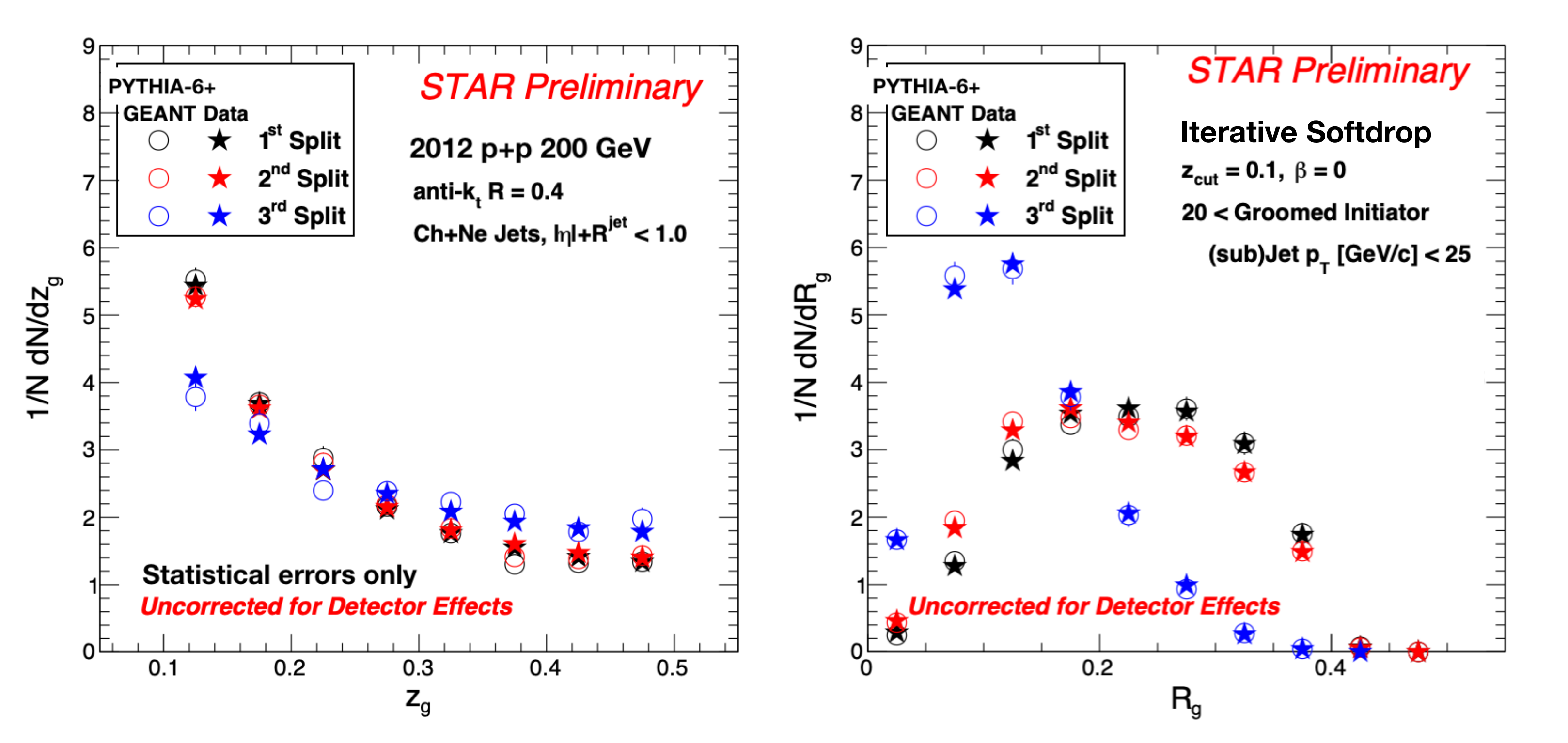} 
   \caption{Measurement of the iterative \zg (left) and \rg (right) in p+p collisions at $\sqrt{s} = 200$ GeV for anti-kt R=0.4 Charged + Neutral jets at different splitting. Each split is selected to have its groomed initiator prong to have a transverse momenta $20 < p_{T} < 25$ GeV. The first, second and third splits is shown in black, red and blue stars for the uncorrected data and compared with simulation in the open circles}
   \label{fig:ppiterzgrg}
\end{figure}

The softdrop algorithm relies on an angular ordered clustered tree and in the previous section we presented the momentum and angular scale measurements at the first split. In applying the softdrop criterion iteratively~\cite{itersoftdrop} along the harder branch in the clustering tree, we are able to qualitatively reconstruct the splitting mechanics during jet clustering. Since the splits along the harder branch are ordered in their opening angle (by construction due to the angular ordering), a formation time argument can be invoked with representing the further splits as also occurring later in time during the parton shower. Such a measurement has the opportunity to experimentally reconstruct a parton shower in formation time and thus have the ability to look at snapshots in time coinciding with the individual splits. This feature is highly desirable in a tomographic study of the QGP since the jet as a whole is demonstrably modified and any experimental evidence of parton shower modifications point to medium properties that vary in formation time. STAR has taken the first step towards measuring the splitting mechanics, via the softdrop observables $z_{g}$ and $R_{g}$ iteratively for the first (black), second (red) and third (blue) splits with their initiator prong's $20 < p_{T} < 25$ GeV as shown in Fig~\ref{fig:ppiterzgrg}. We observe a trend towards a flatter or more symmetric splitting at the third splits as compared to the first splits. These third splits are also much narrower as compared to the first and second splits. These distributions in data (star markers) are not fully corrected for detector effects and they are compared to PYTHIA-6+GEANT simulation (open cricles) which are able to reproduce the distributions as seen in data. 

\section{Jet Angular Scale in Au+Au Collisions}

The Au+Au data used in these proceedings were collected during the 2007 run with its corresponding reference p+p run in 2006 at $\sqrt{s_{\mathrm{NN}}} = 200$ GeV. Since the jet patch trigger is saturated in an Au+Au event, we employ a high tower (HT) trigger, requiring at least one BEMC tower with $E_{T}$ > 5.4 GeV. Event centrality in Au+Au is determined by the raw charged track multiplicity in the TPC within $|\eta| < 0.5$ and we show only events in the 0-20\% centrality range. In Au+Au events, we have two separate jet collections given by the HardCore selection~\cite{starprl}, where jets are clustered with objects (tracks/towers) with $p_{T}> 2$ GeV/$c$, and Matched jets which are clustered from the constituent-subtracted~\cite{cs} event with our nominal $p_{T} > 0.2$ GeV/$c$ for constituents, and are geometrically matched to the HardCore jets ($\Delta R < 0.4$). Further event selection criteria include a minimum \pt~requirement for HardCore di-jets ($p^{\rm{Lead}}_{T} > 16, p^{\rm{SubLead}}_{T} > 8$ GeV/$c$) and an azimuthal angle ($|\Delta \phi (\rm{Lead, SubLead})| > 2\pi/3$) selection to focus on back-to-back di-jets. 


In our studies, we found the groomed jet radii (R$_{g}$) to be highly sensitive to the fluctuating underlying event in Au+Au collisions and therefore we devised a new observable involving sub-jets of a smaller radius reconstructed within the original jet (see here~\cite{SubJetliliana} for a recent theoretical article demonstrating similar classes of observables).
For our nominal anti-$k_{t}$ jets of $R=0.4$, we reconstruct an inclusive set of anti-$k_{t}$ sub-jets with $R=0.1$ from the original jet's constituents. An absolute minimum sub-jet \pt~requirement of $2.97$ GeV/$c$ is enforced in central Au+Au collisions to reduce sensitivity to the background fluctuations. The two observables related to the momentum and angular scales are then defined as follows  $z_{SJ} = \frac{min(p^{SJ1}_{T}, p^{SJ2}_{T})}{p^{SJ1}_{T} + p^{SJ2}_{T}}$, and $\theta_{SJ} = \Delta R (SJ1, SJ2)$,
where $SJ1, SJ2$ are the leading and sub-leading sub-jets, respectively. 

\begin{figure}[h] 
   \centering
   \includegraphics[width=0.8\textwidth]{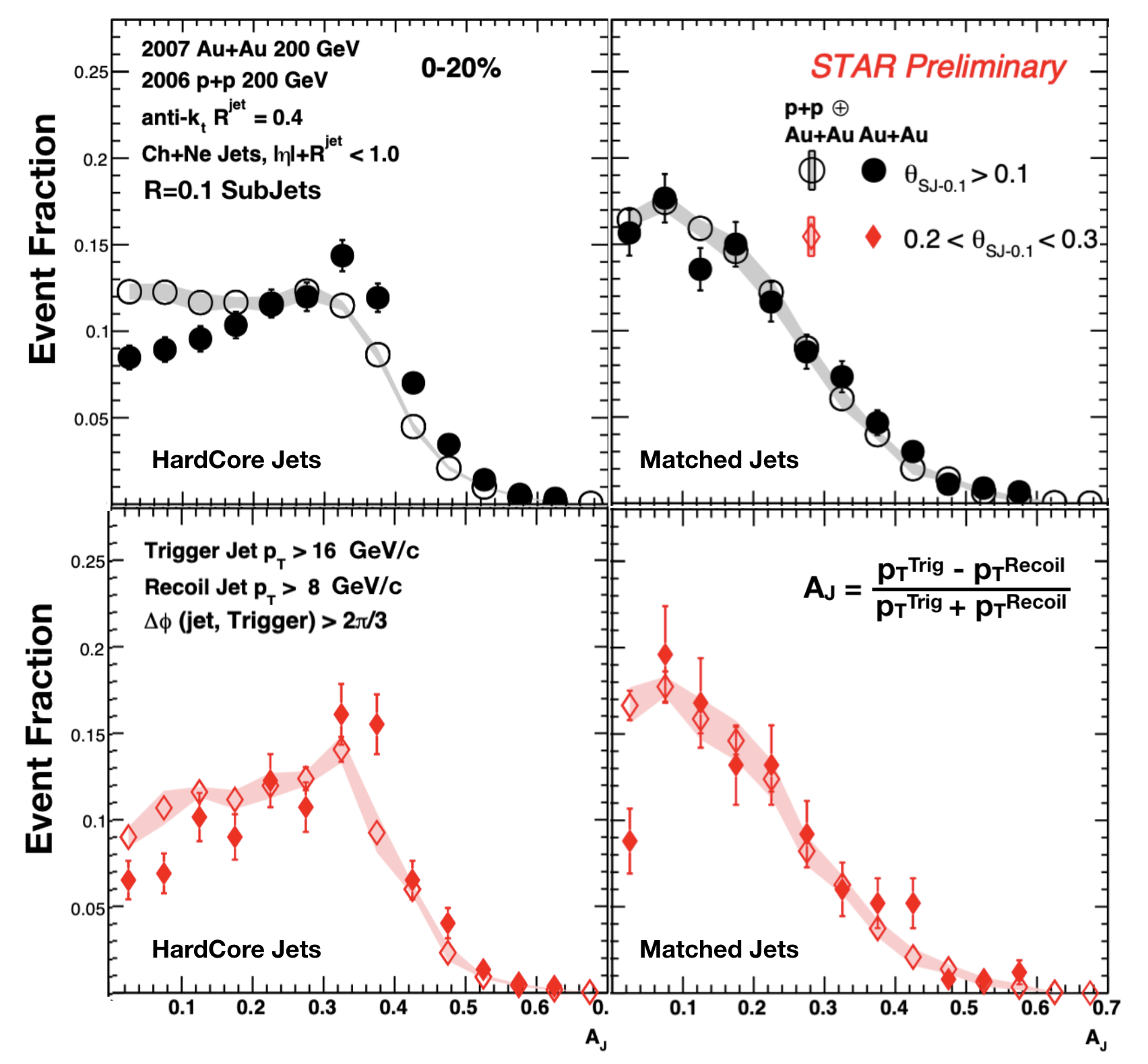} 
   \caption{HardCore and Matched (left figure) di-jet asymmetry ($|A_{J}|$) and the Matched recoil jet yield (right figure) along with the ratios shown in the bottom right panels. Markers are described in the text. 
   }
   \label{fig:aj}
\end{figure}

\begin{figure}[h] 
   \centering
   \includegraphics[width=0.6\textwidth]{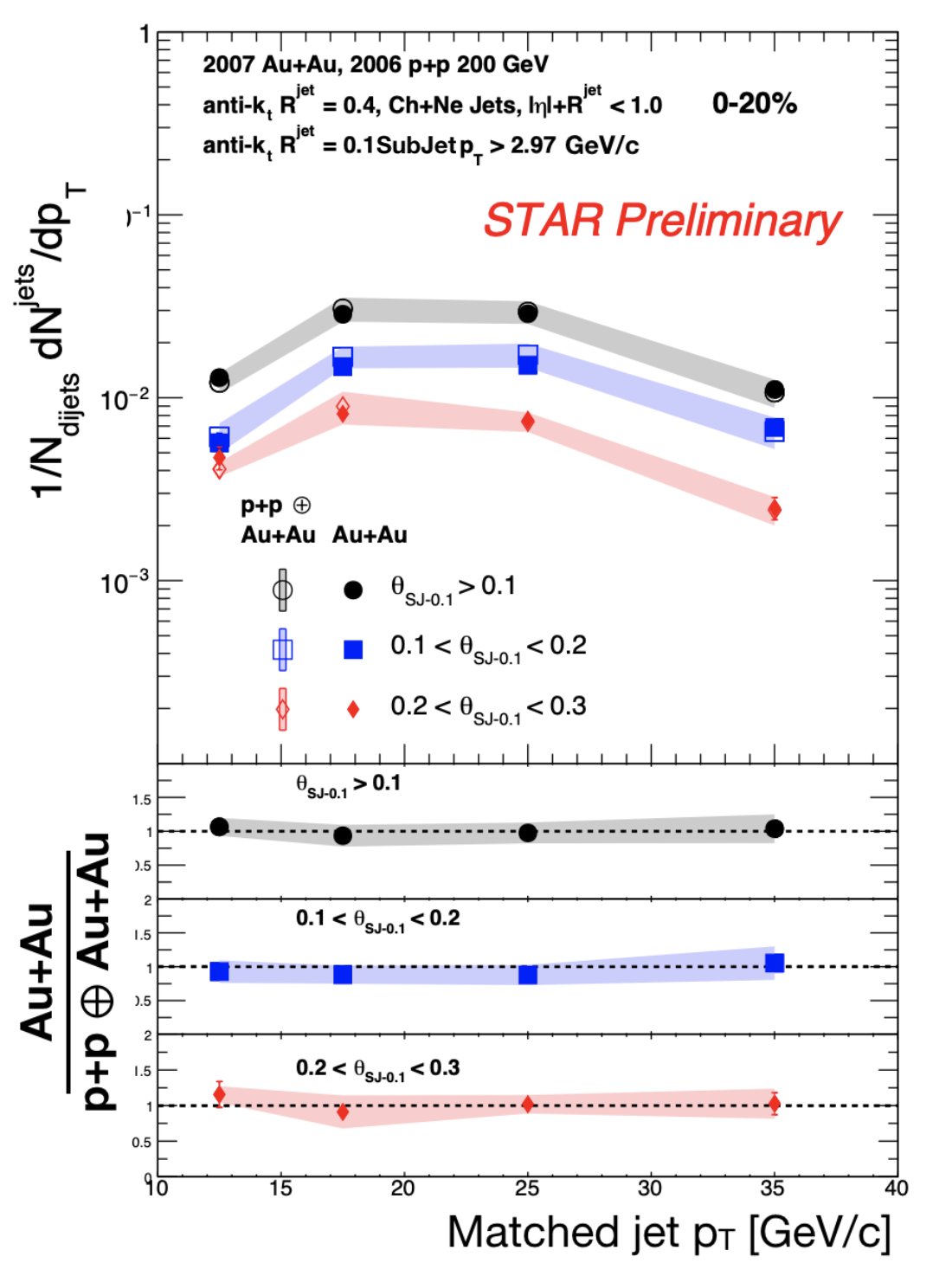} 
   \caption{HardCore and Matched (left figure) di-jet asymmetry ($|A_{J}|$) and the Matched recoil jet yield (right figure) along with the ratios shown in the bottom right panels. Markers are described in the text. 
   }
   \label{fig:aj}
\end{figure}

For a meaningful comparison between Au+Au and a p+p reference, the effects of background fluctuations and detector inefficiencies must be taken into account. To achieve this, HT-triggered p+p data from 2006 is embedded into minimum bias Au+Au data (p+p$\oplus$Au+Au) from 2007, in the same centrality range (0-20\%). During embedding, we account for the relative tracking efficiency (90\%$\pm$7\%) and relative tower energy scale (100\%$\pm$2\%), with a one sigma variation taken as systematic uncertainties.
The TwoSubJet $z_{SJ}$ and $\theta_{SJ}$ distributions for constituent-subtracted Matched recoil jets with $R=0.1$ sub-jets ($SJ-0.1$) recoiling off the trigger (selected with a $|\Delta \phi(\rm{jet, HT})| > 2\pi/3$), in the \pt~range 10-20 GeV/$c$ are observed to be similar in both Au+Au and p+p$\oplus$Au+Au~\cite{raghavHP18}. We also observe a remarkable difference in the shape of $z_{SJ}$ when compared to that of the SoftDrop $z_{g}$, 
which is caused by selecting the core of the jet. The $\theta_{SJ}$ for jets within the considered $p_{T}$ range peaks at small values and includes a natural lower cutoff at the sub-jet radius and we now select jets based on this distribution. 

Di-jet asymmetry for both HardCore and Matched jets (left panels) are shown in Fig.~\ref{fig:aj}. The right panels of Fig.~\ref{fig:aj} show the yield of Matched recoil jets normalized per di-jet for the different $\theta_{SJ}$ selections and the ratios of Au+Au/p+p in the bottom panels. The black, blue and red markers represent recoil jets with selections on $\theta_{SJ}$ $[0.1, 0.4], [0.1, 0.2]$ and $[0.2, 0.3]$ for inclusive, narrow and wide jets, respectively. We observe a clear di-jet imbalance indicating jet quenching effects in the $|A_{J}|$ distributions (comparing Au+Au to p+p$\oplus$Au+Au) for all HardCore jets including the wide angle jets. The Matched jets on the other hand are balanced at RHIC energies, as evident by ratios in the bottom right panels consistent with unity. This is consistent with our earlier measurement~\cite{starprl}. We also note that wide angle jets are still balanced indicating no apparent distinction between wide and narrow jets by the medium in our selection. Further detailed differential analyses are required with the high statistics 2014 data set to extract the medium resolution scale or the coherence length and the effect on standard jet quenching observables at RHIC energies. 


\section{Conclusions}
STAR has presented the first fully unfolded jet substructure measurements including the SoftDrop \zg and \rg and jet Mass of inclusive jets with varying transverse momentum in p+p collisions at $\sqrt{s} = $200 GeV. The measurements are overall reproduced by current leading order Monte Carlo event generators for jets in our kinematic acceptance and reflect the momentum dependent narrowing of jet structure. We also presented for the first time the iterative softdrop splitting observables at first, second and third splits. By comparing the $z_{g}$ and $R_{g}$ at various splits, we find that splits that occur further in the clustering tree have a larger $z_{g}$ on average as opposed to those earlier in the tree. These measurements serve as a first look into an experimental measurement of parton shower of a jet at varying snapshots in time which can then be used to study medium properties in time dependent tomography.  Due to the sensitivity of the SoftDrop observables to the Au+Au underlying event, we introduce and measure the TwoSubJet observables, $z_{SJ}$ and $\theta_{SJ}$ for $R=0.1$ anti-$k_{t}$ sub-jets as representing the momentum and angular scales of a jet in a heavy ion environment. We measure the di-jet momentum asymmetry and the recoil jet yield with the special di-jet selection at STAR and find that HardCore di-jets are imbalanced and Matched di-jets are balanced for jets of varying angular scales. We find no significant difference in the quenching phenomenon for both wide and narrow jets leading to the conclusion that these special jets do not undergo significantly different jet-medium interactions due to their varying angular scales.

\end{document}